\documentclass{mn2e}
\usepackage{amsmath}
\usepackage{textcomp}
\usepackage{amssymb}
\usepackage[pdftex]{graphicx}
\usepackage{amssymb}

\usepackage[margin=.8in, paperwidth=8.5in, paperheight=11in]{geometry}
\usepackage{multicol}
\usepackage{multirow}
\usepackage{epsfig}
\usepackage{graphicx}
\usepackage{dblfloatfix}
\usepackage{accents}
\usepackage{bigdelim}
\newcommand*{\dt}[1]{\accentset{\mbox{\Large\bfseries .}}{#1}}

\begin{document}

\title{Magnetic Field Amplification via Protostellar Disc Dynamos}
\author[S. Dyda et al.]
{\parbox{\textwidth}{S.~Dyda,$^{1,2}$
R.V.E.~Lovelace,$^{3}$
G.V.~Ustyugova,$^{4}$
A.V.~Koldoba$^{5}$ \& I. Wasserman$^{2,3}$\vspace{0.3cm}}\\
 $^{1}$Department of Physics \& Astronomy, University of Nevada Las Vegas, Las Vegas, NV 89154:email: sdyda@physics.unlv.edu\\
 $^{2}$Department of Physics, Cornell University, Ithaca, NY 14853\\
 $^{3}$Department of Astronomy, Cornell University, Ithaca, NY 14853\\
$^{4}$Keldysh Institute for Applied Mathematics, Moscow, Russia\\
$^{5}$Moscow Institute of Physics \& Technology, Dolgoprudny, Moscow Region 141700, Russia
}

\date{\today}
\pagerange{\pageref{firstpage}--\pageref{lastpage}}
\pubyear{2017}

\label{firstpage}

\maketitle

\begin{abstract}

We numerically investigate the generation of a magnetic field in a protostellar disc via an $\alpha \Omega$-dynamo and the resulting magnetohydrodynamic (MHD) driven outflows. We find that for small values of the dimensionless dynamo parameter $\alpha_d$ the poloidal field grows exponentially at a rate $\sigma \propto \Omega_K \sqrt{\alpha_d}$, before saturating to a value $\propto \sqrt{\alpha_d}$. The dynamo excites dipole and octupole modes, but quadrupole modes are suppressed, because of the symmetries of the seed field. Initial seed fields too weak to launch MHD outflows are found to grow sufficiently to launch winds with observationally relevant mass fluxes of order $10^{-9} M_{\odot}/\rm{yr}$ for T Tauri stars. This suggests $\alpha \Omega$-dynamos may be responsible for generating magnetic fields strong enough to launch observed outflows. 

\end{abstract}


\section{Introduction}
Astrophysical jets and winds are a viable mechanism for transporting angular momentum away from protostellar systems and magnetic fields may play a central role in the launching and collimation of these outflows. However, the origin of the large-scale, ordered, magnetic fields required for magnetically driving jets and winds is at present not well understood. One possibility is that the magnetic field is advected inwards from the interstellar medium or through ambipolar diffusion (see Shu, Adams \& Lizano 1987 for review). Another possibility is that the field is generated in the disc via dynamo action (see Brandenburg \& Subramanian 2005 for review).

Magnetic fields play a key role in transporting angular momentum in astrophysical discs, allowing matter to accrete on time scales consistent with models of protostellar evolution. Turbulent viscosity in the accretion disc, most likely generated by the magnetorotational instability (MRI), plays a key role in setting the accretion rate (Balbus \& Hawley 1991). It has been shown that the MRI can amplify and sustain magnetic fields in accretion discs, despite dissipative effects. This has been shown both in shearing box simulations (see for example Brandenburg et al. 1995; Hawley, Gammie \& Balbus 1996; Gressel 2010; Guan and Gammie 2011) and also in global 3D disc simulations (Armitage 1998; Hawley 2000; Hawley 2001; de Villiers \& Hawley 2003).  Magnetic field growth can be modeled as an $\alpha \Omega-$dynamo, where disc rotation generates toroidal field and the $\alpha-$effect (see for example Moffat 1980) converts toroidal field into poloidal field. An important question is whether this dynamo action can generate large-scale fields that are strong enough to magnetically drive outflows.   

Many authors have studied magnetically driven outflows by a priori assuming the system has a large-scale, ordered, magnetic field. Comparatively few studies have investigated magnetohydrodynamic (MHD) driven outflows where the field is generated self-consistently in the disc via dynamo processes. Theoretical work by Campbell (1999, 2003) showed that an $\alpha \Omega-$dynamo operating in the disc changes the wind structure, decreasing the critical launch angle in the classic Blandord-Payne (1982) scenario. Bardou et al. (2001) simulated the back-reaction of outflows on the disc dynamo and found a vertical velocity can enhance the action of the dynamo. Numerical simulations by von Rekowski et al (2003) found that an $\alpha^2$-dynamo can launch  magnetic and thermally driven outflows by growing an initially weak toroidal field. von Rekowski \& Brandernburg (2004, hereafter vRB04) followed this up by showing a stellar dipole threading a dynamo active disc can grow, and undergo episodic accretion/ejection. Stepanovs, Fendt \& Sheikhnezami (2014, hereafter SFS14) showed that episodic outflows can also be generated by having a time dependent disc dynamo. Sadowski et al (2015) modeled a mean-field dynamo around non-spining and spining black holes and found that outflows were more efficient at extracting energy from the system than in thin disc models.      

Observations suggest that dynamo mechanisms may be responsible for the magnetic fields near protostars. Rotational modulation of Zeeman signature in GQ Lup (Donati et al. 2012) and DN Tau (Donati et al. 2013) at different epochs show the magnetic fields of these T Tauri stars evolve in time, which the authors speculate may be due to dynamo processes. In a sample of pre-main sequence stars Vidotto et al (2014) showed that the fields measured from Zeeman broadening (sensitive to large and small scale field) and Zeeman-Doppler imaging (sensitive to only large scale fields) are coupled, suggesting small and large scale fields share the same dynamo generation process. This may have important consequences for outflows, since MHD driven winds are highly sensitive to the geometry of the magnetic field. For instance, Dyda et al (2015, hereafter Dyda15) found that in T Tauri systems disc winds are highly suppressed because of inefficient mass loading if the stellar dipole and disc field are anti-aligned as in vRB04.  

In this paper we perform axisymmetric MHD simulations of an $\alpha \Omega$-dynamo operating inside a protostellar accretion disc. We provide an initial weak poloidal seed magnetic field and track its growth and the corresponding MHD outflows. Our goal is to undertand how the disc dynamo affects the corresponding outflows and the late time magnetic field structure. The structure of this paper is as follows. In Section 2 we review the disc dynamo contribution to the equations of motion and describe our MHD code. In section 3 we present results, describing how outflows and magnetic field structure around the star are affected by the initial seed field and the strength of the dynamo, the disc viscosity and the magnetic diffusivity. We conclude in Section 4 and discuss implications for future work and observational prospects for classical T Tauri stars (CTTS).

\section{Dynamo Model}
We investigate the roles of disc dynamos by performing 2D axisymmetric simulations using a Godunov-type MHD code (Koldoba et al. 2015). We describe the basic equations solved, including the addition of a $\alpha \Omega$-dynamo module in Section \ref{sec:basic_eq}. The evolution of the field in the limit of a thin disc, the ``no-z" approximation, is discussed in Section \ref{sec:disc_average}. We describe the initial conditions in Section \ref{sec:IC} and the boundary conditions in Section \ref{sec:BC}. Our simulations are performed in dimensionless units, and we introduce appropriate units for T Tauri stars in Section \ref{sec:dimensions}.
        
\subsection{Basic Equations}
\label{sec:basic_eq}
Using the prescription of mean-field dynamo theory (see for example Kulsrud 1999), fields are broken up into a mean, large-scale component (denoted by an overbar) and a turbulent, stochastic component (denoted by a $\delta)$. The velocity and magnetic field are respectively
\begin{subequations}
 \begin{equation}
\mathbf{v} = \overline{\mathbf{v}} + \delta \mathbf{v},
\end{equation}
\begin{equation}
\mathbf{B} = \overline{\mathbf{B}} + \delta \mathbf{B}.
\end{equation} 
\end{subequations} 
The plasma flows are assumed to be described by the equations of non-relativistic MHD in a non-rotating reference frame
\begin{subequations}
 \begin{equation}
\frac{\partial \rho}{\partial t} + \nabla \cdot \left( \rho \overline{\mathbf{v}} \right) = 0~, 
\end{equation}
\begin{equation}
 \frac{\partial \rho \overline{\mathbf{v}}}{\partial t} + \nabla \cdot \mathcal{T} = \rho \mathbf{g}~,
\end{equation}
\begin{equation}
\frac{\partial \overline{\mathbf{B}}}{\partial t} = \nabla \times (\overline{\mathbf{v}} \times \overline{\mathbf{B}}) + \nabla \times \overline{\mathcal{E}} - \eta \nabla \times ( \nabla \times \overline{\mathbf{B}}) ,
\label{eq:induction}
\end{equation}
\begin{equation}
\frac{\partial \left( \rho S \right)}{\partial t} + \nabla \cdot \left( \rho  \overline{\mathbf{v}}S \right) = \mathcal{Q}~,
\end{equation}
\end{subequations}
where $\rho$ is the mass density, $S$ is the specific entropy, $\mathcal{T}$ is the momentum 
flux density tensor, $\mathcal{Q}$ is the rate of change of entropy per unit volume due to heating, $\eta$ is the magnetic diffusivity and $c$ is the speed of light. The mean electromotive force
\begin{equation}
\overline{\mathcal{E}} = \overline{\delta \mathbf{v} \times \delta \mathbf{B}},
\end{equation}
is generated by the turbulent fluctuations of the velocity and magnetic field. Assuming fluctuations are isotropic and uncorrelated 
\begin{equation}
\overline{\mathcal{E}} = \alpha \overline{\mathbf{B}} - \eta_t \nabla \times \overline{\mathbf{B}},
\label{eq:EMF}
\end{equation}
where $\alpha = -\tau/3 \overline{\left( \delta \mathbf{v} \cdot \nabla \times \delta \mathbf{v} \right)}$
is the product of the decorrelation time of the velocity fluctuations $\tau$ and the mean helicity of the turbulence and the turbulent magnetic diffusivity $\eta_t = 1/3 \tau \overline{\delta \mathbf{v}^2}$. We parametrize the strength of the mean electromotive force in our simulations in terms of dimensionless $\alpha$ parameters 
\begin{equation}
\overline{\mathcal{E}} = \alpha_d Z \Omega_K \overline{\mathbf{B}} + \alpha_{\eta} \frac{c_s^2}{\Omega_K} \nabla \times \overline{\mathbf{B}}  ~,
\label{eq:dynamo_term}
\end{equation}
where $\alpha_d$ and $\alpha_{\eta}$ parametrize the strength of the dynamo and magnetic diffusivity respectively. The accompanying prefactors, the disc height $Z$, local Keplerian angular velocity $\Omega_K$ and midplane sound speed $c_s$, are chosen by dimensional analysis.  As detailed at the end of this section,  $\alpha_d$ and $\alpha_{\eta}$ are non-zero only inside the disc.  In axisymmetry, magnetic diffusion is the only mechanism to suppress the dynamo and prevent the gas from becoming overmagnetized. If $\alpha_d$ is too large then the assumption that the fluctuations grow more slowly and thus remain subdominant to the mean-field breaks down and our paramterization of the electromotive force is no longer justified. 

We define the dynamo number 
\begin{equation}
N_d = \frac{\alpha_d \Omega_K h^3}{\eta_t^2} \approx \frac{\alpha_d}{\alpha_{\eta}^2},
\end{equation}
where h is the half-thickness of the disc. Dynamo effects are thought to be important for $N_d > N_c$ for some critical dynamo number (Stepinski \& Levy 1988).    

We assume that heating is offset by radiative cooling so that $\mathcal{Q}=0$. Also, $\mathbf{g} = -\left(GM/r^2 \right)\hat{r}$ is the gravitational acceleration due to the central mass $M$. We model the fluid as a non-relativistic ideal gas with equation of state
\begin{equation}
 S = \ln\left( \frac{p}{\rho^{\gamma}}\right)~,
\end{equation}
where $p$ is the pressure and $\gamma = 5/3$. 

We use cylindrical coordinates $(R,\phi,Z)$ as these are the coordinates used by our code. Sometimes it will be useful to also refer to spherical coordinates $(r,\theta,\phi)$ as well. 

We assume the magnetic diffusivity is primarily due to turbulence i.e $\eta_t \gg \eta$ and is parametrized by (\ref{eq:EMF}). Likewise we assume that kinematic viscosity is due to turbulent fluctuations of velocity and magnetic field and use the $\alpha$-disc parametrization (Shakura and Sunyaev 1973) 
\begin{equation}
 \nu_t = \alpha_{\nu} \frac{c_s^2}{\Omega_K}~,
\end{equation}
where $\alpha_{\nu}\leq 1$ is a dimensionless constant. The ratio, 
\begin{equation}
 \mathcal{P} = \frac{\alpha_{\nu}}{\alpha_{\eta}}~ ,
\end{equation}
is the magnetic Prandtl number of the turbulence in the disc. We take $\alpha_{\nu} = \alpha_{\eta} = 0.1$ unless otherwise stated.

The momentum flux density tensor is given by
\begin{equation}
 \mathcal{T}_{ik} =p\delta_{ik}+ \rho \overline{v}_i \overline{v}_k + \left( \frac{\overline{\mathbf{B}}^2}{8 \pi} \delta_{ik} - \frac{\overline{B}_i \overline{B}_k}{4 \pi} \right) + \tau_{ik}~,
\end{equation}
where the bracketed term, the so-called Maxwell stress, is responsible for the Lorentz force and $\tau_{ik}$ is the viscous stress contribution from the turbulent fluctuations of the velocity. 
    As mentioned, we assume that these can be represented in the same
way as the collisional viscosity by substitution of the turbulent viscosity. The leading order contributions to the viscous stress arise from large velocity gradients. In a Keplerian type disc these are dominated by the azimuthal terms $v_{\phi} \sim v_{K}$. However, we include all viscous terms involving $v_R$ and $v_{\phi}$.
   The leading order contribution to the momentum flux density from 
turbulence are therefore
\begin{align}
\tau_{R\phi} = -\nu_t \rho R \frac{\partial \overline{\Omega}}{\partial R} ~, \hspace{1.2cm}
\tau_{Z \phi} = - \nu_t \rho R \frac{\partial \overline{\Omega}}{\partial z}~, \nonumber \\
 \tau_{Z R} = - \nu_t \rho \frac{\partial \overline{v}_R}{\partial Z}~, \hspace{1.2cm}
\tau_{R R} = - 2 \nu_t \rho \frac{\partial \overline{v}_R}{\partial R}~, \nonumber \\
 \tau_{\phi \phi} = - 2 \nu_t \rho \frac{\overline{v}_R}{R} ~, \hspace{4cm}
\end{align}
where $\overline{\Omega} = \overline{v}_{\phi}/R$ is the angular velocity of the gas.

    The transition from the viscous/diffusive disc to the 
    non-viscous/non-diffusive corona is handled  by multiplying the alpha 
coefficients ($\alpha_\nu$ and $\alpha_\eta$)  by a dimensionless factor $\xi(\rho)$ which varies smoothly from $\xi=1$ for $\rho \geq \rho_d=0.75\rho(R,Z=0)$   to
$\xi =0$ for $\rho\leq 0.25 \rho_d$ as described in Appendix B
of Lii, Romanova, \& Lovelace (2012).  The dynamo coefficient is smoothly 
reduced to zero at the inner radius $R_{\rm{dyn}} = 5$ and in the corona via
\begin{equation}
  \alpha(R,Z) =
  \begin{cases}
   0 & R<R_{\rm{dyn}}~, \\
   \alpha_{d} Z \sqrt{\frac{GM}{R^3}} e^{-(\rho_{\rm{dyn}}/\rho)^2} \tanh(\frac{R-R_{\rm{dyn}}}{\Delta R}) & R  >R_{\rm{dyn}}~.
  \end{cases}
\label{eq:dynamocoefficient}
\end{equation}
where $\rho_{\rm{dyn}} = 0.1$ and $\Delta R = 1$. We found it necessary to impose an inner disc cutoff to avoid numerical instabilities at the surface of the star. Since our code does not treat the subgrid physics of the turbulent velocity and magnetic field, in the remainder of this paper we will supress the overbar and take $\mathbf{v}$ and $\mathbf{B}$ to be the mean, large-scale velocity and magnetic field. 

\subsection{Disc Averaged Equations}
\label{sec:disc_average}
To gain some insight into the dynamo process, consider a thin, axisymmetric, Keplerian disc with diffusivity $\eta$ and radial infall velocity $v_R$ operating an $\alpha \Omega$-dynamo. We will ignore any vertical motion $v_Z$ and neglect the $Z$ dependence of $\Omega$ and $v_R$, often refered to as the ``no-z" approach (see Phillips 2001 for a review). Dynamical quantities refer to the mean, large-scale components and we reserve the over-bar to refer to the vertically averaged quantity. Breaking the induction equation (\ref{eq:induction}) into (R, $\phi$, Z) components, we find
\begin{subequations}
\begin{equation}
\begin{split}
-\frac{1}{R}\frac{\partial}{\partial Z}\left( \frac{\partial \Psi}{\partial t} \right) =& 
                                 \frac{\partial}{\partial Z}\Bigg[ \Bigg. -\alpha_d Z \Omega_K B_{\phi} + \frac{v_R}{R}\frac{\partial \Psi}{\partial R}
                                \\ &- \frac{\eta}{R}\left( R \frac{\partial}{\partial R} \left(\frac{1}{R}\frac{\partial \Psi}{\partial R} \right)        + \frac{\partial^2 \Psi}{\partial Z^2}\right)\Bigg. \Bigg],
\end{split} 
\end{equation}
\begin{equation}
\begin{split}
\frac{\partial B_{\phi}}{\partial t} = -& \frac{\partial \Omega}{\partial R} \frac{\partial \Psi}{\partial Z} 
+ \frac{\partial}{\partial Z}\left( - \frac{\alpha_d Z \Omega_K}{R}\frac{\partial \Psi}{\partial Z} + \eta \frac{\partial B_{\phi}}{\partial Z}  \right)  \\
&- \frac{\partial}{\partial R} \left( \frac{\alpha_d Z \Omega_K}{R}\frac{\partial \Psi}{\partial R} + v_R B_{\phi} - \frac{\eta}{R}  \frac{\partial (R B_{\phi})}{\partial R} \right),  
\end{split}
\end{equation}
\begin{equation}
\begin{split}
\frac{\partial}{\partial R}\left( \frac{\partial \Psi }{\partial t} \right) =& \frac{\partial}{\partial R}\Bigg[ \Bigg. \alpha_d Z \Omega_K B_{\phi} R - v_R \frac{\partial \Psi}{\partial R} \\ &+ \eta \left( R \frac{\partial}{\partial R} \left(\frac{1}{R}\frac{\partial \Psi}{\partial R}\right) + \frac{\partial^2 \Psi}{\partial Z^2} \right)\Bigg. \Bigg],
\end{split}
\end{equation}
\label{eq:thin_disc}
\end{subequations}
where $\Psi$ is the magnetic flux function. Two types of solutions are possible, depending on the symmetries of $\Psi$ and $B_{\phi}$ - a ``dipole" like solution where $\Psi(Z) = \Psi(-Z)$ and $B_{\phi}(Z) = -B_{\phi}(-Z)$ and a ``quadrupole" solution where $\Psi(Z) = -\Psi(-Z)$ and $B_{\phi}(Z) = B_{\phi}(-Z)$ (see Kulsrud 2005). To solve the problem exactly we make a gauge choice, for example $\Psi(0) = 0$ and boundary conditions at $Z = \pm h$ such as $\partial \Psi / \partial Z = 0$. Intuitively then, one may think of the flux function as a cosine in the dipole like case and as a sine in the quadrupole case
\begin{equation}
  \Psi(Z) =
  \begin{cases}
   \Psi_0 \cos \left( \frac{\pi Z}{h} \right) & ``\rm{dipole}", \\
   \Psi_0 \sin \left( \frac{\pi Z}{2h} \right) & ``\rm{quadrupole}".
  \end{cases}
\label{eq:dynamo_psi}
\end{equation}
We work in the thin disc, $h \ll R$, approximation. The variation in the flux function over the half disc thickness is therefore in each case $\Delta \Psi \approx \Psi_0$ but the variation in its derivative for each case is $\Delta \Psi / \Delta Z _{\rm{dip}} \approx 2 \Delta \Psi / \Delta Z _{\rm{quad}}$.  
Therefore integrating over the upper half part of the disc, and replacing all quantities with their vertically averaged quantities $\int_0^h B_{\phi} dz = h \bar{B}_{\phi}$, $\int_0^h -\frac{1}{R}\frac{\partial \Psi}{\partial Z} = h \bar{B}_R$ and ignoring terms involving $B_Z$, the R and $\phi$ components give us the coupled system 
\begin{subequations}
\begin{equation}
\frac{\partial \bar{B}_{\phi}}{\partial t} = (\frac{3}{2} C_1 + C_2 \alpha_d) \Omega \bar{B}_R - \frac{\eta}{h^2}\bar{B}_{\phi},
\end{equation}
\begin{equation}
\frac{\partial \bar{B}_{R}}{\partial t} = C_2 \alpha_d \Omega \bar{B}_{\phi} - \frac{\eta}{h^2}\bar{B}_R.
\end{equation}
\label{eq:integrated_induction}
\end{subequations}
where $C_1 = 1 (-1)$ and $C_2 = 2 (-1)$ in the dipole (quadrupole) case.
Using $\eta = \alpha_{\eta} \Omega h^2$, we combine (\ref{eq:integrated_induction}) and find
\begin{equation}
\begin{split}
\frac{\partial^2 \bar{B}_{R}}{\partial t^2} &+ 2 \alpha_{\eta} \Omega \frac{\partial \bar{B}_{R}}{\partial t} \\ &+ \left[ \alpha_{\eta}^2 - C_2 \alpha_d \left( 3/2 C_1 + C_2 \alpha_d \right) +  \right] \Omega^2 \bar{B}_{R} = 0. 
\end{split}
\end{equation} 
Taking $\bar{B}_{R} \propto \exp(\sigma \Omega t)$ we find growth rates 
\begin{equation}
\sigma  = - \alpha_{\eta} \pm \sqrt{C_2 \alpha_d (3/2 C_1 + C_2 \alpha_d)}.
\label{eq:growth}
\end{equation}
Taking $\alpha_d \ll 1$ we approximate
\begin{equation}
  \sigma  + \alpha_{\eta} \approx
  \begin{cases}
   \pm \sqrt{3 \alpha_d}  & ``\rm{dipole}", \\
   \pm \sqrt{3 \alpha_d /2} & ``\rm{quadrupole}".
  \end{cases}
\label{eq:growth_t}
\end{equation}
Substituting back into (\ref{eq:integrated_induction}) we find
\begin{equation}
  \bar{B}_R \approx
  \begin{cases}
   \pm \sqrt{\frac{2 \alpha_d}{3}} \bar{B}_{\phi}  & ``\rm{dipole}" , \\
   \mp \sqrt{\frac{2 \alpha_d}{3}} \bar{B}_{\phi} & ``\rm{quadrupole}".
  \end{cases}
\label{eq:bfield}
\end{equation}
The toroidal field component is primarily generated by the differential rotation of the disc. It is also the dominant contribution to the magnetic field. Therefore, at late times when the matter and magnetic pressure in the disc are in equilibrium, we expect the toroidal component to saturate. Equation (\ref{eq:bfield}) implies that the radial field will also saturate, to a value proportional to $\sqrt{\alpha_d}$. For this reason, we scale all our magnetic field plots to $\sqrt{\alpha_d}$ (see Section \ref{sec:stellar_multipoles}). Equation (\ref{eq:growth_t}) shows that for $\alpha_d,\alpha_{\eta} \ll 1$ the growth rate $\sigma \propto \sqrt{\alpha_d}$ so we scale all times by a factor of $\alpha_d^{-1/2}$ (see Section \ref{sec:growth_rate}). We note that requiring a positive growth rate $\sigma > 0$ implies $N_d \gtrsim 1/3$. Though we have neglected order one numbers, this provides an intuitive picture for why there is a critical dynamo number - if the diffusion time scale is too fast compared to the dynamo time scale then the field diffuses away and never grows. Stepinski \& Levi (1988) found purely growing dynamo modes for dynamo number $R_m = \left( R/h\right)^{3/2} \alpha_d/\alpha_{\eta}^2 \sim 100$ which agrees with this result up to some $\mathcal{O}(1)$ numbers. 

Above we neglected terms that depended on $\partial/\partial R$ to make the exponential growth of the field via the dynamo apparent. However, these terms are important for determining the effects of viscosity and diffusivity in the disc. In particular, taking $v_R = \nu/R$ and $\nu, \eta$ constant and expanding $(\ref{eq:thin_disc})$ yields advection like terms $\propto (-\nu + \eta) \partial \Psi/ \partial R$ and $\propto (-\nu - \eta) \partial B_{\phi}/ \partial R$. $\eta$ and $\nu$ may cooperate or compete in transporting different field components. We show this empirically in our simulations where the effects of the dynamo is changed by varying the Prandtl number $\mathcal{P} = \alpha_{\nu}/\alpha_{\eta}$ (see Section \ref{sec:diffusivity}).

\subsection{Initial Conditions}
\label{sec:IC}

\subsubsection{Magnetic Field \& Dynamo}
Our simulations use a disc field (Zanni, 2007) as a seed field for the dynamo, defined by the poloidal flux function
\begin{equation}
 \Psi_{\rm{disc}} = \frac{4}{3} B_0 R_0^2 \left( \frac{R}{R_0} \right)^{3/4} \frac{m^{5/4}}{\left( m^2 + Z^2/R^2\right)^{5/8}},
\end{equation}
where the poloidal magnetic field components can be computed via
\begin{equation}
 B_Z = \frac{1}{R} \frac{\partial \Psi}{\partial R}, \hspace{2cm} B_R = - \frac{1}{R} \frac{\partial \Psi}{\partial Z}. 
\end{equation}
The parameter $m$ determines how much the field lines bend in the R-Z plane with the limit $m \rightarrow \infty$ corresponding to purely vertical
field lines. In this study we set $m = 1$.

The field generated by the dynamo can be expanded using a multipole expansion
\begin{equation}
\Psi = \mu_{\rm{dip}} \Psi_{\rm{dip}} + \mu_{\rm{quad}} \Psi_{\rm{quad}} + \mu_{\rm{oct}} \Psi_{\rm{oct}} + ...
\label{eq:multipole}
\end{equation}
where 
\begin{subequations}
\begin{equation}
\Psi_{\rm{dip}} = \frac{R^2}{(R^2+Z^2)^{\frac{3}{2}}}, 
\end{equation}
\begin{equation}
\Psi_{\rm{quad}} = \frac{3}{4} \frac{R^2 Z}{\left(R^2 + Z^2\right)^{5/2}},  
\end{equation}
\begin{equation}
\Psi_{\rm{oct}} = \frac{R^2(4Z^2 - R^2)}{2(R^2+Z^2)^{7/2}},
\end{equation}
\end{subequations}
are the dipole, quadrupole and octupole contributions respectively. These are expected to be the dominant modes excited by the disc dynamo. We show in Appendix \ref{sec:multipoles} how we numerically extract the multipole moments from our simulation. 

\subsubsection{Matter Distribution}
Initially the matter of the disc and corona are assumed to be in mechanical equilibrium (Romanova et al. 2002). 
The initial density distribution is taken to be barotropic with
\begin{equation}
  \rho(p) =
  \begin{cases}
   p/T_{\rm{disc}} & p>p_b ~~{ \rm and} ~~ R \geq R_d~, \\
   p/T_{\rm{cor}} & p<p_b ~~ {\rm or} ~~ R  \leq R_d~,
  \end{cases}
\end{equation}
where $p_b$ is the level surface of pressure that separates the cold matter of the disc from the hot matter of the corona and $R_d$ is the initial inner disc radius.   At this surface
the density has an initial step discontinuity from value $p/T_{\rm{disc}}$ to $p/T_{\rm{cor}}$.

Because the density distribution is barotropic, the initial angular velocity is a constant on coaxial cylindrical surfaces about the $Z-$axis. Consequently, the pressure 
can  be determined from the Bernoulli  equation
\begin{equation}
 F(p) + \Phi + \Phi_c = \mathcal{B}_0~,
\label{eq:forcebalance}
\end{equation}
where $\mathcal{B}_0 = 3 \times 10^{-4}$ is Bernoulli's constant, $\Phi = -GM/\sqrt{R^2 + Z^2}$ is the gravitational potential, with $GM = 1$ in the code, $\Phi_c = \int_{R}^{\infty} r dr ~\omega^2(r)$ is the 
centrifugal potential, and
\begin{equation}
  F(p) =
  \begin{cases}
   T_{\rm{disc}}\ln(p/p_b) & p>p_b ~~ {\rm and} ~~ R \geq R_d~, \\
   T_{\rm{cor}}\ln(p/p_b) & p<p_b ~~{\rm or} ~~ R \leq R_d~.
  \end{cases}
\end{equation}
The initial half thickness of the disc $h/R = 0.1$ at the initial inner disc edge $R_d = 5$.

\subsubsection{Angular Velocity}
     The initial angular velocity of the disc is slightly sub-Keplerian, 
\begin{equation}
 \Omega = (1-0.003)\Omega_K(R) \hspace{1cm} R>R_d~.
\end{equation}
Inside of $R_d$, the matter rotates rigidly with angular velocity
\begin{equation}
 \Omega = (1-0.003)\Omega_K(R_d) \hspace{1cm} R  \leq R_d.
\end{equation}
The corotation radius $R_{c}=(GM/\Omega_*^2)^{1/3} = 2$ is the radius where
the angular velocity of the disc equals that of the star.

\subsection{Boundary Conditions}
\label{sec:BC}

We follow as closely as possible the treatment in Dyda15. Our simulation uses a grid $N_R\times N_Z =154 \times 230$ cells. The star is taken to be cylindrical in shape with radius $R_* = 1$ and height $Z_* = 2$, extending 10 grid cells above and below the disc midplane. In the R-direction, the first 60 grid cells have length $dR = 0.05$. Later cell lengths are given recursively by $dR_{i+1} = 1.025 dR_{i}$. Similarly, in the Z-direction the first 30 grid cells above and below the equatorial plane have length $dZ = 0.05$. Later cell lengths are given recursively by $dZ_{j+1} = 1.025 dZ_{j}$.

We assume axisymmetry about the axis. On the star and the external boundaries we want to allow fluxes and impose free boundary conditions $\partial \mathcal{F}/\partial n = 0$ where $\mathcal{F}$ is a dynamical variable and $n$ is the vector normal to the boundary. At the external boundary we impose outflow boundary conditions. We also require that along the edge of the disc matter be inflowing $v_R \leq 0$ and in the corona matter be outflowing $v_r \geq 0$. On the star we require that matter on the stellar boundary flow into the star $v_r < 0$. We treat the corner of the star by averaging over the nearest neighbour cells in the R and Z directions.

\subsection{Dimensional Variables}
\label{sec:dimensions}
\begin{table}
\begin{center}
  \begin{tabular}{ | l  c  l |}
                                                                 \\\hline \hline
Parameters               & Symbol                 & Value                      \\ \hline \hline
   mass                  & $M_0$                  & $1.59\times10^{33}$ g       \\
   length                & $R_0$                  & $1.50\times10^{12}$cm        \\
   magnetic field        & $B_0$                  & $8.04\times10^{-2}$ G                 \\ \hline 
   time                  & $P_0$                  & $3.53\times10^{-2}$ y     \\
   velocity              & $v_0$                  & $8.42\times10^{7}$cm/s    \\ 
   density               & $\rho_0$               & $7.24\times 10^{-15} $ g/cm$^3$\\
   accretion rate        & $\dt{M}_0$             & $2.72\times 10^{-7}$ $M_\odot$/yr \\
   temperature           & $T_0$                  & $4.26\times 10^{7}$ K \\
   dipole strength       &$\mu_{\rm{dip}}$        & $2.69\times 10^{35}$ $\rm{G \ cm}^3$\\
   quadrupole strength   &$\mu_{\rm{quad}}$       & $4.04\times 10^{47}$ $\rm{G \ cm}^4$\\
   octupole strength     &$\mu_{\rm{oct}}$        & $6.05\times 10^{59}$ $\rm{G cm}^5$\\
        \hline \hline
  \end{tabular}
\end{center}
\caption{Mass, length, and magnetic field scales of interest and the corresponding scales of other derived quantities for a CTTS. One can obtain the physical values from the simulation values by multiplying by the corresponding dimensional quantity above.}
\label{table:units} 
\end{table}

The MHD equations are written in dimensionless form so that the simulation results can be applied  to different types of stars. We assume that the central object is a CTTS with mass $M_* = 0.8 M_{\odot}$, a radius $R_* = 2R_{\odot}$ and a magnetic field with magnitude $B_* = 3 \times 10^3$ G on the stellar surface, which is typical for the magnitude of the stellar dipole. We define dimensionful quantities with a 0 subscript, to denote typical values of physical parameters at a reference radius $R_0$. The mass scale is set by the stellar mass so $M_0 = M_*$. The reference length, $R_0 = 0.1 \rm{AU}$, is taken to be the scale of a typical inner disc radius. Assuming a stellar dipole field, the magnetic field strength $B_0 = B_* (R_*/R_0)^3$. The mass, length and magnetic field scales allow us to define all other dimensionful quantities.

The reference value for the velocity is the Keplerian velocity at the radius $R_0$, $v_0 = (GM_0/R_0)^{1/2}$. The reference time-scale is the period of rotation at $R_0$, $P_0 = 2\pi R_0/v_0$. From the relation $\rho_0 v_0^2 = B_0^2/4 \pi$, we define the reference density $\rho_0$ of the disc. The reference mass accretion rate is $\dt{M}_0 = 4 \pi \rho_0 v_0 R_0^2$. The reference temperature $T_0 = v_0^2/2 \times m_H/k_B$ where $m_H$ is the atomic mass of hydrogen and $k_B$, the Boltzmann constant, is obtained by taking the ratio of the reference pressure $p_0 = B_0^2/8 \pi$ and the reference density. The dimensions of the coefficients in the multipole expansion of the flux function are obtained by scaling the reference field by the appropriate power of $R_0$ - $\mu_{\rm{dip}} = B_0 R^3_0$, $\mu_{\rm{quad}} = \mu_{\rm{dip}} R_0$ and $\mu_{\rm{oct}} = \mu_{\rm{dip}} R_0^2$.  

Results obtained in dimensionless form can be applied to objects with widely different sizes and masses. However, the present work focuses on CTTS with the typical values shown in Table \ref{table:units}. One can obtain dimensionful quantities from simulation results by multiplying by the appropriate dimensionful quantity above.

\section{Results}

\begin{figure*}
                \centering
                \includegraphics[width=0.95\textwidth]{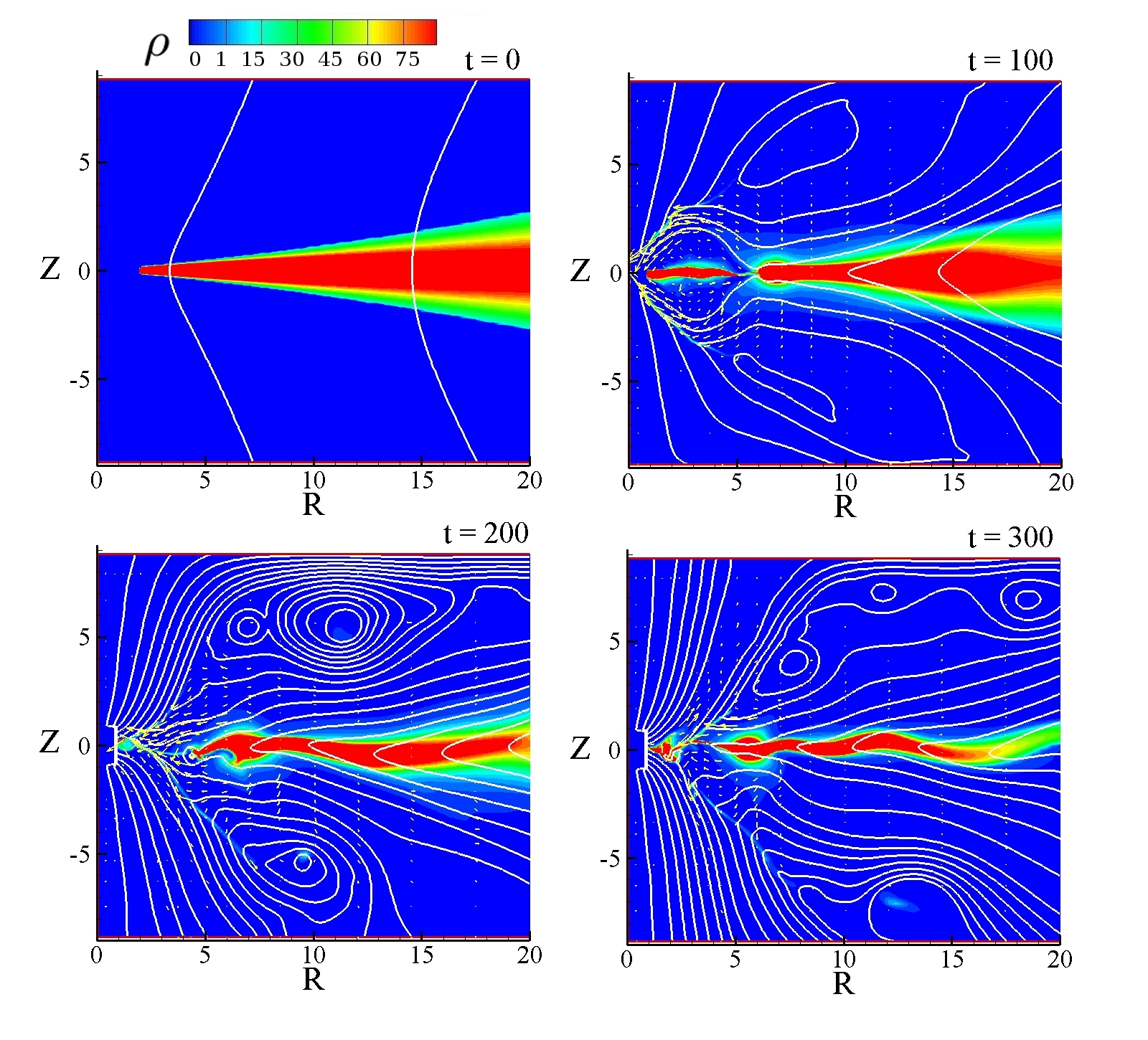}
        \caption{Density $\rho$ (color), poloidal field lines $\Psi$ (white) and poloidal velocity $v_{\rm{p}}$ (yellow) for the case $N_d = 10$. t = 0 shows the initial large-scale field. By t = 100 the dynamo has changed the field structure in the inner part of the disc. By t = 200 the field has developed an parallel dipole structure near the star. By t = 300 the dipole structure is maintained on the star and a quadrupole structure begins to form near the inner disc with field lines being parallel to the disc, rather than threading through it. The outer parts of the disc have not yet evolved this quadrupolar structure.  
}
\label{fig:poloidal_weak}
\end{figure*} 

We provide an initially weak, poloidal, seed field threading the disc. Differential disc rotation causes this poloidal field to twist, producing a toroidal component. The $\alpha \Omega$-dynamo in turn converts this toroidal component into additional poloidal field. This is a positive feedback loop and the poloidal field continues to grow until other physical effects such as diffusion halts the growth.  

In Fig. \ref{fig:poloidal_weak} we plot disc density $\rho$ (color), magnetic flux contours $\Psi$ (white) and poloidal velocity vectors $v_{\rm{p}}$ (yellow) of our fiducial $\alpha_d = 0.1$ run. We see that magnetic flux in the disc increases as a function of time and the flux builds up on the star as it advects inwards with the accreting matter. On time scales of the simulation, the dynamo only changes the field geometry in the inner part of the disc $R \lesssim 10$ where loops form. Along the star, a diple-like structure forms, as flux from the disc grows and is advected onto it. For $t  \gtrsim 200$ a quadrupole structure forms near the inner disc with field lines \emph{parallel} to the disc, rather than threading \emph{through} it.   In the outer parts of the disc field lines become more inclined relative to their initial configuration due to magnetic pressure in the corona, as in the case where the dynamo is not operating, shown in Dyda15. Despite the dynamo operating for all radii $R > R_{\rm{dyn}} = 5$, the simulation does not run long enough for the outer parts of the disc to evolve this quadrupolar structure. This requires diffusing the dipole-like flux lines threading through the disc out the simulation domain, which occurs in the outer disc on timescales $\sim (\alpha_{\eta} \Omega_K )^{-1} \approx 900$ inner disc orbits.

Runs were performed with seed field amplitudes ranging between $B_0 = 10^{-2} - 10^0$. These correspond to midplane plasma $\beta = 10^1 - 10^5$. For $\beta \gg 10^1$, corresponding to seed fields $B_0 \ll 1$, the results were largely independent of the initial magnetic field. For larger field values, the dynamo had comparatively little effect, and the simulation behaved as for the case $\alpha_d = 0$ where the disc dynamo is explicitly turned off (see Dyda et al 2015). The dipole and higher order multipoles grow on the star as flux is generated in the disc and advected onto the star. We note that Cowling's antidynamo theorem (1934), which states that a self-sustained, axisymmetric field cannot be maintained, does not apply to our simulations because the dynamo term in (\ref{eq:dynamo_term}) models subgrid physics which is by assumption non-axisymmetric.     

\subsection{Stellar Multipoles} 
\label{sec:stellar_multipoles}

\begin{figure}
                \centering
                \includegraphics[width=0.45\textwidth]{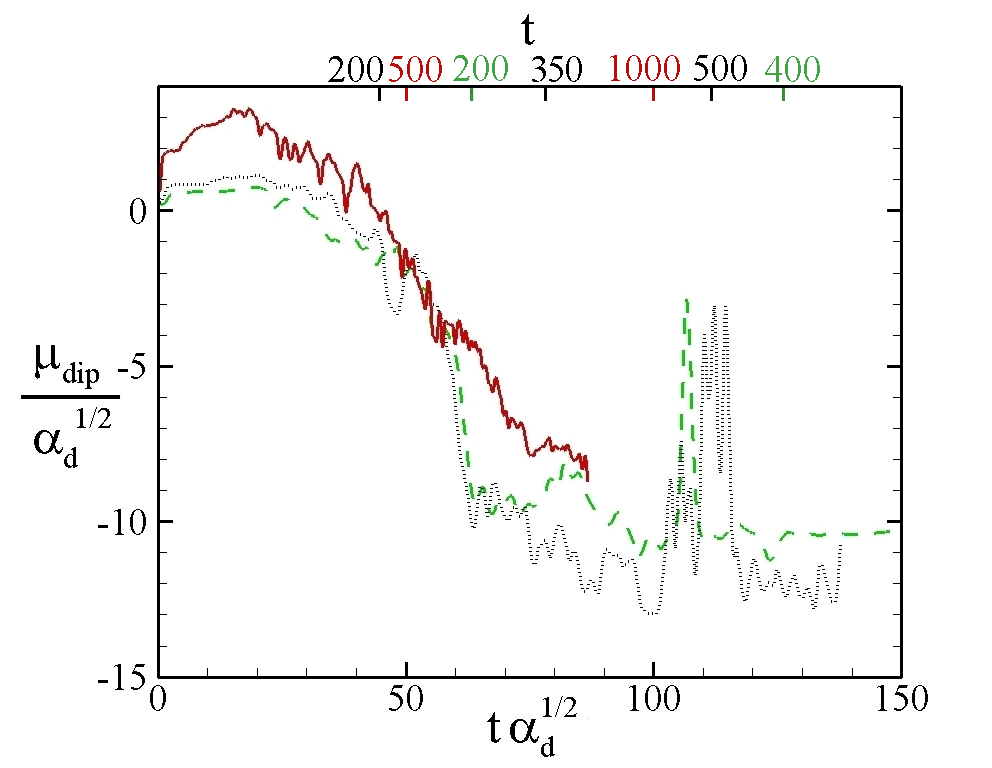}
                \includegraphics[width=0.45\textwidth]{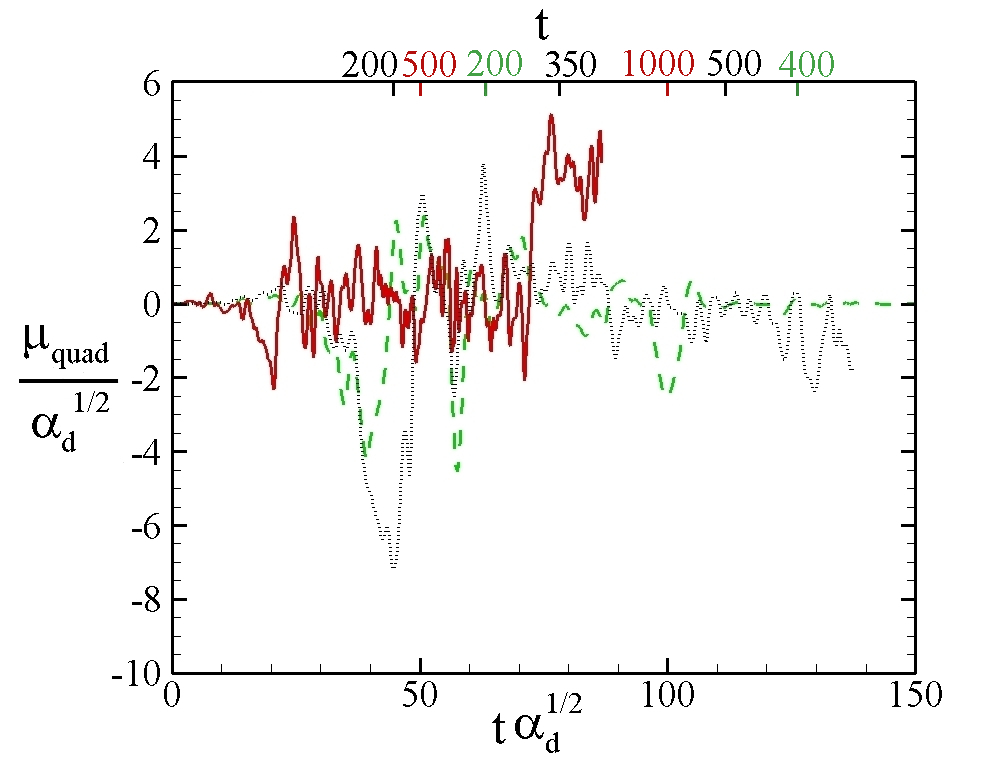}
                \includegraphics[width=0.45\textwidth]{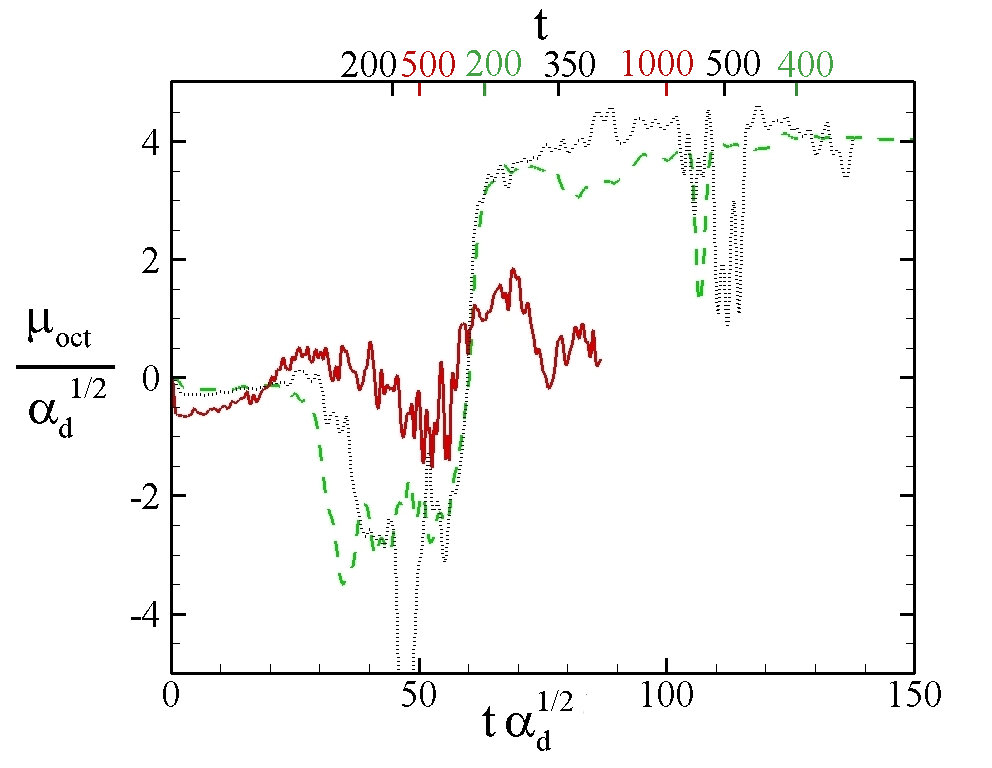}
        \caption{Stellar multipole moments as a function of time for dynamo numbers $N_d = 1$ (red, solid), 5 (black, dot) and 10 (green, dash) with constant disc diffusivity $\alpha_{\eta} = 0.1$. We see the multipole moments saturate to a value that scales with $\alpha_d^{-1/2}$ and of the growth rate scale to $\alpha_d^{1/2}$.  \textit{Top}  Dipole moment $\mu_{\rm{dip}}$ is the dominant contribution to the field \textit{Middle} Quadrupole moment $\mu_{\rm{quad}}$ oscillates around zero \textit{Bottom} Octupole moment $\mu_{\rm{oct}}$ saturates to a value of the opposite sign to the dipole moment
}
\label{fig:dip_dyn_scaled}
\end{figure}

Any magnetic field can be expanded in a general multipole expansion (\ref{eq:multipole}). We show how we extract the stellar dipole $\mu_{\rm{dip}}$, quadrupole $\mu_{\rm{quad}}$ and octupole $\mu_{\rm{oct}}$ moments from our simulations in Appendix \ref{sec:multipoles}.  We use this method to study the time dependence of the magnetic field topology around the star.

In Fig \ref{fig:dip_dyn_scaled} we plot  the dipole (top panel), quadrupole (middle panel) and octupole (bottom panel) moments  as a function of time for dynamo numbers $N_d = $ 1 (red, solid), 5 (black, dot) and 10 (green, dash) (corresponding to $\alpha_d =$ 0.01, 0.05 and 0.1 respectively) for fixed diffusivity $\alpha_{\eta} = 0.1$ and seed field amplitude $B_0 = 0.1$. We have scaled the time by $\alpha_d^{1/2}$ and the dipole moment by $\alpha_d^{-1/2}$, as suggested by the linear analysis of Section \ref{sec:disc_average}. Empirically, we see this scaling yields similar behaviour for runs with different dynamo strength. We find qualitatively different behaviour of the dipole, quadrupole and octupole moments. The dipole moment grows to a value of approximately $\mu_{\rm{dip}} \approx - N_d^{1/2}$. It saturates to this value on time scales $\tau \sim \alpha_d^{-1/2} \Omega_K(R_{\rm{dyn}})^{-1}$. The quadrupole moment oscillates about $\mu_{\rm{quad}} = 0$, which is what we expect given the symmetry of the initial seed field. The magnitude of these oscillations increases with increasing dynamo number. At late times the amplitude of the oscillations dies down. The octupole moment begins to oscillate at the same time that the dipole mode begins to grow. In the cases where $\alpha_d = 0.05 \ \rm{or} \ 0.1$, when the dipole reaches its asymptotic value, the octupole moment quickly assumes the opposite sign so that the late time octupole moment saturates at the same time as the dipole but with opposite polarity. This suggests that the lower order multipole moments are excited first, and higher order multipoles are excited and grow as the lower order multipoles saturate. The octupole moment saturates to a value of $\mu_{\rm{oct}} \approx - N_d^{1/2}$ as suggested by the linear theory. For weaker dynamos, the higher order modes are not excited.  These results are consistent with the symmetries of the seed field - the disc dynamo is a symmetric function of position Z, therefore we expect to excite symmetric (dipole and octupole) and not asymmetric (quadrupole) modes.  Reversing the polarity of the seed field reverses the signs of the generated multipoles, but does not otherwise change the results, as expected. 

The field dynamics is largely determined by the physics in the inner disc. In particular, we performed simulations where we varied the cutoff radius $5 \leq R_{\rm{dyn}} \leq 8$ but kept the dynamo coefficient $\alpha$ in (\ref{eq:dynamocoefficient}) constant at $R_{\rm{dyn}}$ by varying $\alpha_d$. The late time moments changed by roughly 10\% and growth rates and times were similarly affected. Poloidal field plots retained the same qualitative character, where a large loop formed near the cutoff radius but field lines in the outer disc tilted more towards the equatorial plane of the disc. 

Though the disc dynamo generates a predominantly dipole component of the magnetic field, it is unclear whether a disc dynamo can generate a closed magnetosphere which has been shown to generate strong outflows in the propelor regime. Rather this may require a dynamo mechanism operating on the star itself.

\subsection{Growth Rate}
\label{sec:growth_rate}

\begin{figure}
                \centering
                \includegraphics[width=0.45\textwidth]{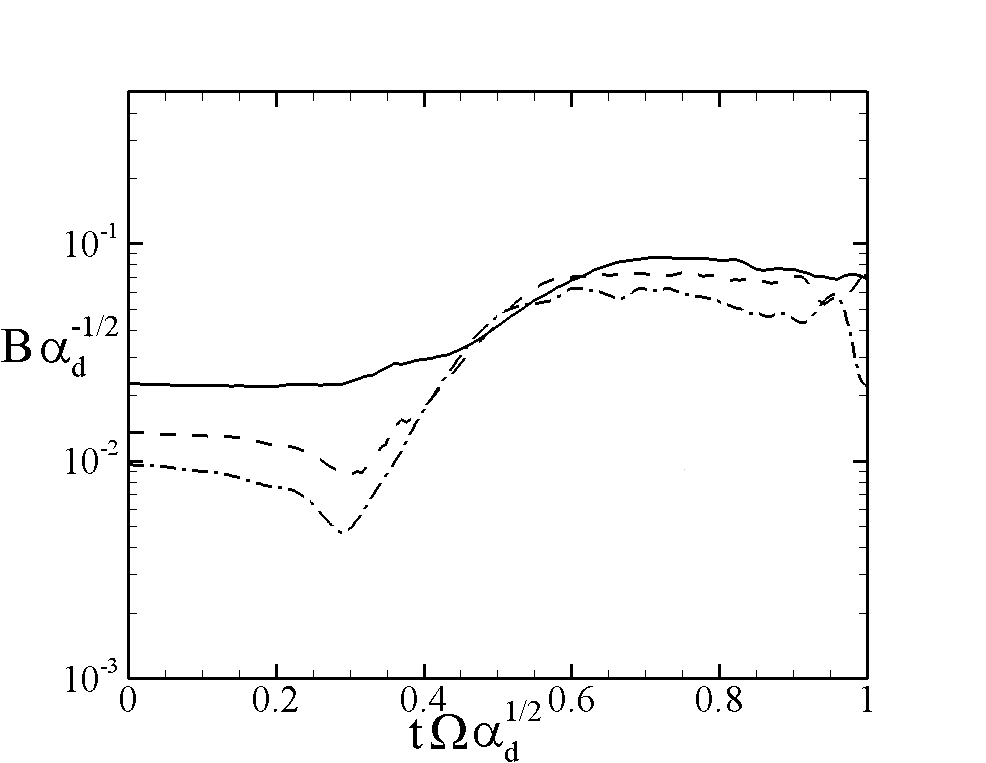}
        \caption{Vertically integrated magnetic field $\bar{B}_Z$ at radii $R = 10$ (solid), 15 (dash) and 20 (dash-dot) as a function of angular phase $\Omega t$ scaled by $\alpha_d^{1/2}$ for the case $\alpha_d = 0.1$. The exponential growth predicted by the linearized regime occurs for phase angles $t \Omega \sqrt{\alpha_d} \ll 1$  
}
\label{fig:bz_vs_t_scaled}
\end{figure} 

The disc averaged equations (\ref{eq:growth_t}) predict that the exponential growth rate $\sigma \propto \sqrt{\alpha_d}~ \Omega_K$, scaling like $\sqrt{\alpha_d}$ and the local Keplerian angular velocity. In Fig \ref{fig:bz_vs_t_scaled} we plot vertically integrated $\bar{B}_Z$ at radii $R = 10, 15 \ \rm{and} \ 20$ as a function of angular phase $\Omega_K t$ for the case $\alpha_d = 0.1$. We scale the magnetic field by $\alpha_d^{-1/2}$ and the phase by $\alpha_d^{1/2}$ for ease in comparison of the other simulations. We see that there is indeed a period of exponential growth lasting a time $\sim 1$ where the field changes by a factor $\sim 10$. The scaling (\ref{eq:growth_t}) suggests that $\sigma \approx \sqrt{3}$ in these units but we find, averaging the rates for these three radii and $\alpha_d = 0.01, 0.05 \ \rm{and} \ 0.1$ runs $\sigma = 2.0 \pm 1.0$. This suggests that our linear analysis, where we dropped order one numbers, is valid during this short time.

\begin{figure}
                \centering
                \includegraphics[width=0.45\textwidth]{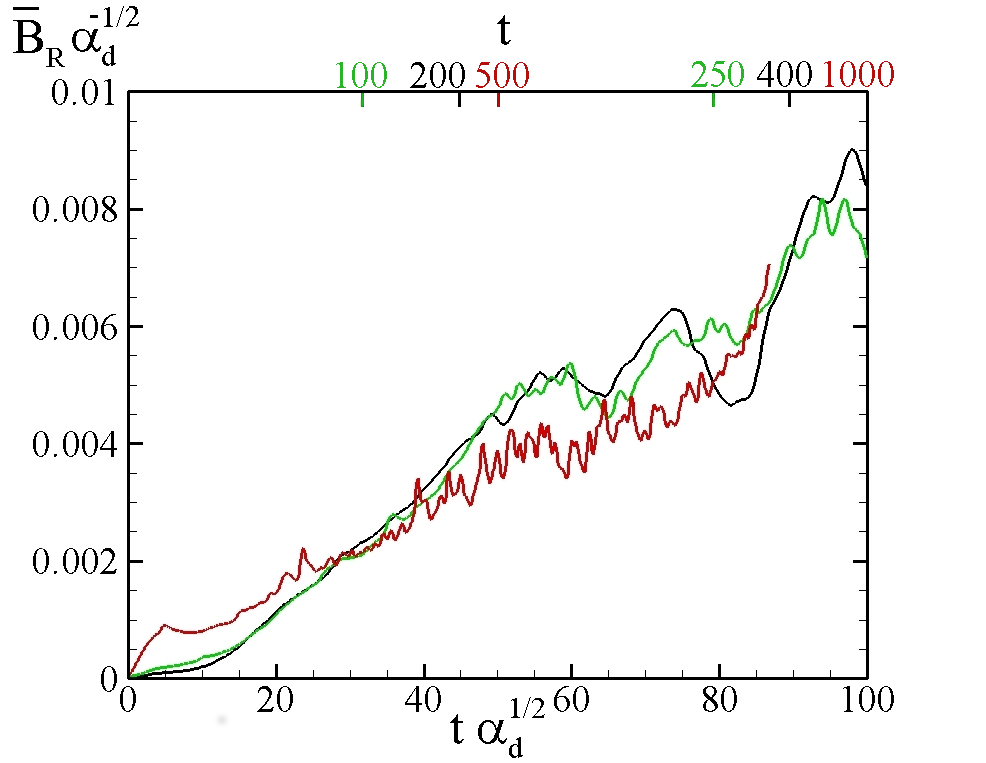}
                \includegraphics[width=0.45\textwidth]{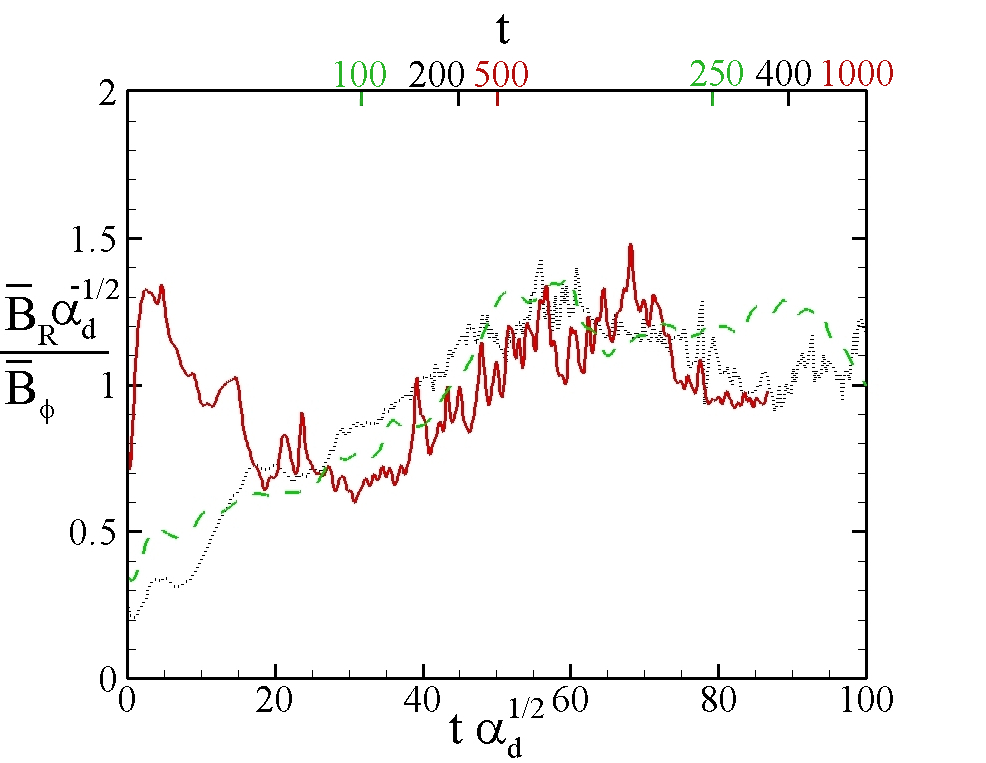}
        \caption{Average magnetic field in the disc computed from equation (\ref{eq:bbar_int}) for the cases $\alpha_d$ = 0.01 (red), 0.05 (green) and 0.1 (black). \textit{Top -} Average radial magnetic field strength $\bar{B}_R$.  Unlike the field around the star, the disc averaged magnetic field does not have time to saturate on time scales of the simulation. \textit{Bottom -} Ratio of average radial to toroidal magnetic field $\bar{B}_R/\bar{B}_{\phi}$. This ratio varies by an order one number, suggesting that toroidal field is built up via differential rotation and a fraction is converted to poloidal field via the $\alpha-$effect.}
\label{fig:br_disc}
\end{figure}

The magnetic field strength in the disc is also found to increase in time. We define the averaged field in the disc via
\begin{equation}
\bar{B_i} = \left(\frac{\int_{\rho > \rho_{\rm{floor}}} (B_i)^2 \ dV}{\int_{\rho > \rho_{\rm{floor}}} \ dV}\right)^{1/2},
\label{eq:bbar_int}
\end{equation}
where $\rho_{\rm{floor}} = 1$. We define the averaged field in this way because $B_R$ and $B_{\phi}$ are antisymmetric about the disc plane and average out to zero vertically. In Fig \ref{fig:br_disc} we plot $\bar{B}_{R}$ (top panel) and $\bar{B}_{R}/\bar{B}_{\phi}$ (bottom panel) as functions of time. We see that the radial magnetic field is an increasing function of time and grows nearly linearly. The ratio of radial to toroidal magnetic field is, to an order one number, constant over the simulation time. This suggests that the simple picture whereby differential rotation creates toroidal field, and the $\alpha \Omega$-dynamo transforms this field into poloidal field is qualitatively correct. This is the distinguishing feature of the $\alpha \Omega-$dynamo, where toroidal field is generated via differential rotation and poloidal field is generated via the $\alpha-$effect. Unlike on the surface of the star, where the field saturates at late times in the simulation, the integrated field in the disc continues to grow. Since the growth rate is set by the local Keplerian time, on time scales of the simulation the outer part of the disc does not have time to saturate with magnetic field, whereas near the star, where time scales are expected to be shorter and field is advected inwards at a faster rate, the field has time to saturate. We can estimate the growth rate of $\bar{B}$ by assuming a local exponential growth rate $\sigma t \Omega_K$ and initially uniform field $B_0$ in which case $\int B_0^2 e^{2\sigma t \sqrt{GM} R^{-3/2}} R dR \sim t^{4/3}$. If the field were decreasing radially then this dependence on t would be weaker. The greatest contribution of this integral occurs for $\sigma \Omega t \gtrsim 1$ so there is a characteristic radius $R_{\rm{max}} \propto t^{2/3}$ beyond which there is little contribution. We find $\bar{B} \sim t^{1.1}$ so a slightly stronger dependence than predicted by this estimate. However, it provides a semi-quantitative understanding of how polynomial growth is achieved globally in the disc whereas growth is locally exponential.     

\subsection{Outflows}
\begin{figure}
                \centering
                \includegraphics[width=0.45\textwidth]{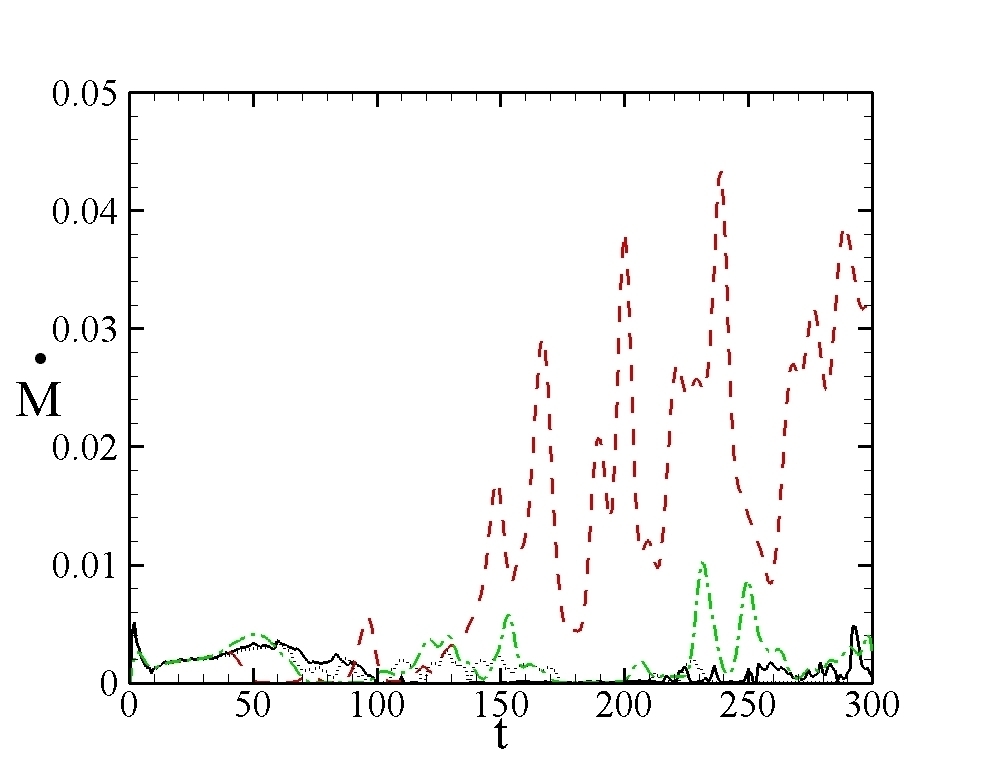}
        \caption{ Total mass flux $\dot{M}$ as a function of time for dynamo numbers $N_d = 0 \ (\rm{black, dotted}), 1 \ (\rm{black, solid}), 5 \ (\rm{green, dash-dot}),10 \ (\rm{red, dash})$ ($\alpha_d$ = 0, 0.01, 0.05, 0.1). Observationally significant outflows occur after t = 150.}
\label{fig:outflow}
\end{figure}

\begin{figure}
                \centering
                \includegraphics[width=0.45\textwidth]{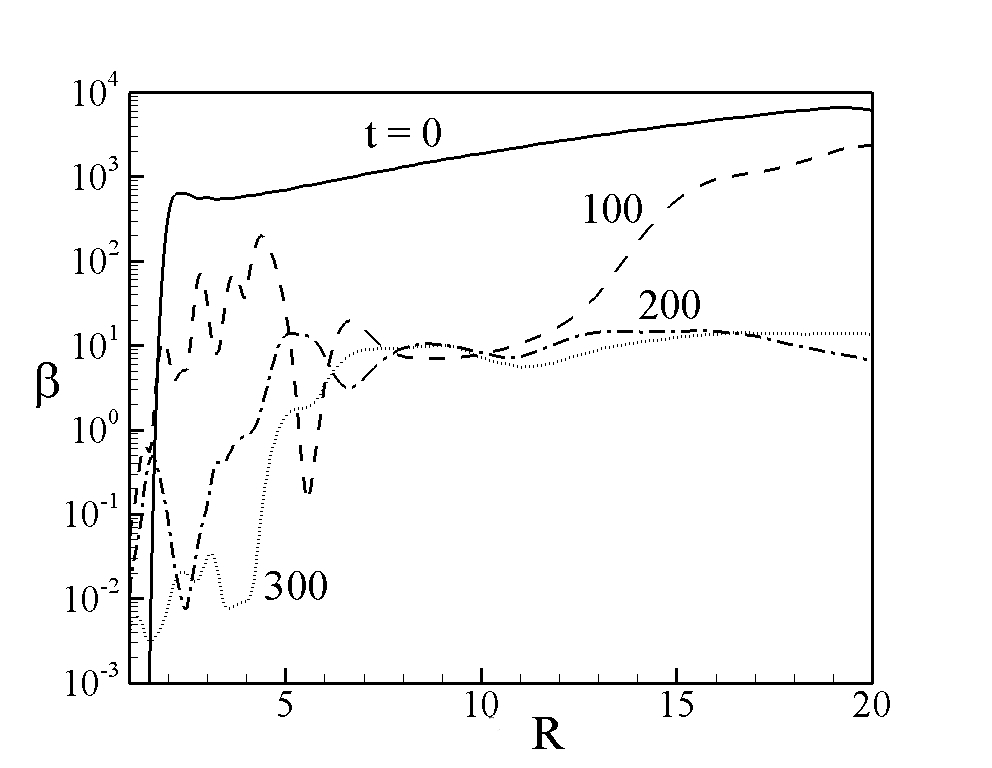}
        \caption{Plasma $\beta = 8 \pi P / |\mathbf{B}|^2$ as a function of radius for t = 0 (solid), 100 (dash), 200 (dash-dot) and 300 (dotted) in the case $\alpha_d = 0.1$. When the inner disc $R < 5$ becomes magnetically dominated, an inner disc wind begins to outflow. 
}
\label{fig:beta_t}
\end{figure}

An important question is how the field generated by the dynamo affects matter outflows. In Fig \ref{fig:outflow} we plot the total mass flux $\dt{M}$ at the outer boundaries $Z = \pm 5$ for various dynamo numbers. Choosing to measure mass flux along boundaries in the range $3 < Z < 8$ had little effect on the results. We also require that any matter flux have a velocity $|v_p| > 0.1$, which corresponds to an observationally relevant $v \sim 200 \rm{km/s}$. We note that non trivial outflows are generated for dynamo number $N_d = 10$. The outflows begin at approximately $t = 150$ inner disc orbits, corresponding to a time $t \sim 5 ~ \rm{yrs}$ for T Tauri stars. This is when the dipole moment reaches its saturation value and the plasma $\beta < 1$ in the inner parts of the disc near $R = R_{\rm{dyn}}$. This timescale is independent of the magnitude of the seed field, since the growth time is set by the dynamo number. Disc dynamos are therefore suitable candidates to explain time varying outflows observed in T Tauri systems with observed time scales between outflows of $10 - 100 \rm{yrs}$. Such a model has previously been explored by SFS14 where the dynamo was explicitly turned on and off to produce time varying outflows.  Though oscillatory, the mass flux $\dt{M} \approx 0.025$ or approximately $6.8 \times 10^{-9} M_{\odot}/\rm{yr}$ for a T Tauri star, is consistent with observed values. The start of outflow generation corresponds to a decrease in the ratio of matter to magnetic pressure $\beta$. In Fig \ref{fig:beta_t} we plot this parameter, averaged over the thickness of the disc, as a function of radius for $t = 0, \ 100, \ 200 \ \rm{and} \ 300$ . Winds are primarily launched from the inner part of the disc. We see that the dynamo free region $R_{*} < R < 5$ becomes magnetically dominated ($\beta < 1$) for $t > 200$, consistent with the formation of a gap for $R < R_{\rm{dyn}}$ and our assumption that the dynamo operates in a matter dominated disc still holds. The time of outflow generation also corresponds to the time at which the stellar multipoles asymptote to their late time values. At this time our linear analysis (\ref{eq:thin_disc}) breaks down because we can no longer ignore the $v_Z$ velocity component. Flux can now flow in the Z direction, thereby halting the growth of flux on the surface of the star.  We note that though the stellar dipole and octupole moments saturate, other features of the disc remain non-stationary - the accretion mass flux varies in time and is the source of time dependence of the wind. This is consistent with the picture that winds are primarily affected by the disc, in particular near the inner disc where the wind mass flux is greatest. Outflows can be enhanced via dynamo growth on time scales of the simulation because the timescales for magnetic field growth are shorter than the diffusion timescale. Compare this with the case $\alpha_d = 0.01$ where now diffusion effects are $\sim 10$ times more important. The field saturates to a value roughly $\sqrt{10}$ times smaller and the outflows $\sim 10$ times weaker.  

\subsection{Dipole Seed Field}
The time when observationally significant outflows are launched corresponds to the time when the stellar dipole moment saturates to $\mu_{\rm{dip}} \approx 3$.   We compare this to a run where the initial magnetic field is a stellar dipole with magnetic moment $\mu_{\rm{dip}} = 3$. We keep other parameters unchanged, $\alpha_d = \alpha_{\eta} = \alpha_{\nu} = 0.1$. This field configuration generates outflows roughly an order of magnitude smaller than the late time dynamo configuration, but at earlier times - since the stellar dipole moment is already present, it launches outflows right away without first having to grow flux in the disc and advect it onto the star. Torkelsson and Brandenburg (1994) showed that if a stellar dipole is used as the seed field, the field generated in the disc has opposite polarity. Dyda15 showed that this type of configuration, where the stellar dipole and disc fields are anti-aligned, has suppressed outflows compared to a pure disc field because the field line structure in the inner part of the disc is not conducive to mass loading. Our results here are consistent with this, where the case with a seed disc field launches the strongest outflows.  

The dipole moment grows and saturates to $\mu_{\rm{dip}} \approx 4$, and the octupole moment saturates to $\mu_{\rm{oct}} \approx - 0.5$. We observe the same behaviour as for the disc field where the dipole and octupole have the same sign until saturation, at which time it quickly reverses direction. For comparison, if the dynamo is turned off, this stellar dipole configuration settles to a configuration with $\mu_{\rm{dip}} \approx 2.5$ (owing to diffusion of flux through the disc) and the higher order moments are zero. This emphasizes that for magnetically driven winds the field geometry, and not only the field strength, plays an important role in setting the mass flux.


\subsection{Magnetic Diffusivity}
\label{sec:diffusivity}
\begin{figure}
                \centering
                \includegraphics[width=0.45\textwidth]{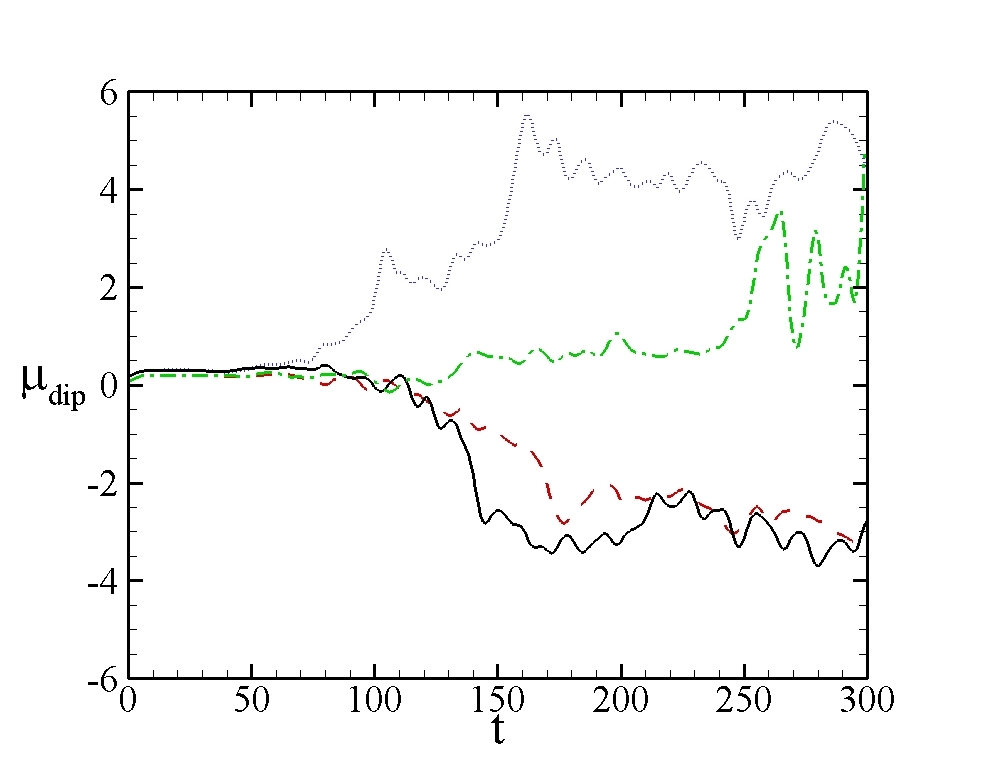}
        \caption{Dipole moment $\mu_{\rm{dip}}$ as a function of time for fixed $\alpha_d = 0.1$ and viscosity $\alpha_{\nu} = 0.3$ with varying magnetic diffusivity $\alpha_{\eta} = $ 0.1 (black, solid), 0.07(red, dash),  0.05 (green, dash-dot) and 0.03 (blue, dotted). 
}
\label{fig:diffusivity}
\end{figure}

We vary the dynamo number by keeping $\alpha_d = 0.1$ fixed and varying the magnetic diffusivity from $0.03 \leq \alpha_{\eta} \leq 0.3$. The viscosity is also kept constant at $\alpha_{\nu} = 0.3$. In Fig \ref{fig:diffusivity} we plot the stellar dipole moment as a function of time for these various runs. As before, the dipole moment saturates to a certain value but interestingly, the sign depends on the magnetic diffusivity. For $\alpha_{\eta} \leq 0.05$ then $\mu_{\rm{dip}} > 0$ whereas for $\alpha_{\eta} \geq 0.07$ then $\mu_{\rm{dip}} < 0$. Empirically the field line structure is different in these two regimes. In the case of larger diffusivity, the field lines in the outer part of the disc are mostly vertical, threading \emph{through} the disc. In contrast, for the smaller diffusivity cases the field lines are nearly \emph{parallel} with the disc. When $\eta$ is smaller, the field is more strongly frozen-in, so as matter accretes in the disc it drags the footpoint inwards and the fieldline bends towards the disc. When $\eta$ is larger, matter can effectively accrete without draging the footpoint and the fieldline remains vertical (Dyda et al 2013). Decreasing the dynamo parameter $\alpha_d = 0.01$ had no effect on the sign of the generated dipole moment - in fact, the late time value $\mu_d \propto \alpha_d^{-1/2}$ holds for $\alpha_d < 0.1$ suggesting that the effect is not due to a changing dynamo number. Rather, the sign of the late time dipole moment depends on the sign of the poloidal field at early times near the star, which is controlled by the relative strengths of viscous and diffusive effects. The Prandtl number in the case where a large positive dipole grew was $\mathcal{P} = 10$ whereas for other runs this ratio was $\mathcal{P} \lesssim 1$. In cases where the late time dipole moment asymptotes to a negative value, the poloidal field in the upper hemisphere is negative at early times. This is reversed for cases where the dipole moment asymptotes to a positive value. As described in Section \ref{sec:disc_average}, changing the relative strength of the viscosity and diffusivity changes the sign of a term governing the flux function evolution. Since the flux function is coupled to the toroidal field, we expect this to change the sign of the toroidal field. 

\section{Discussion}

We found an initially poloidal magnetic field threading the disc grows via the effects of an $\alpha \Omega-$dynamo. The differential rotation of the disc grows a toroidal magnetic field, which is converted to poloidal magnetic field via the $\alpha-$effect. This process repeats until at late times the poloidal field strength saturates to a value $\propto \sqrt{\alpha_d}$ when field growth is countered by diffusion and magnetic pressure prevents additional flux from advecting onto the star. The local growth rate of the field during early times is exponential and set by the local Keplerian velocity and the strength of the dynamo with $\sigma \propto \Omega_K \alpha_d^{1/2}$  

The dynamo primarily excites dipole and octupole modes, allowing for the growth of these modes on the star. Quadrupole modes are suppressed at late times, owing to the symmetry of the initial seed field.  An important finding in our work is dipole and octupole modes saturate to values of opposite polarity with magnitudes $\propto \sqrt{\alpha_d}$.  T Tauri stars have been observed using the Doppler-Zeeman technique to have dipole and higher multipole moments (Donati \& Collier Cameron 1997; Donati et al. 1999; Jardine et al. 2002; Johnstone et al 2014). We may thus be able to distinguish between dynamo models operating on the star and in the disc by observing the relative signs of the stellar multipoles. It is important to accurately model higher order multipoles, as assuming a pure dipole structure leads to overestimating of the size of the magnetosphere and thus accretion/outflow rates. 

A seed field that is initially too weak to generate observable outflows can grow on time scales $\tau \sim \Omega_K ^{-1} \alpha_d^{-1/2}$ to values which are strong enough to magnetically launch winds. This suggests that disc dynamos may be important in generating the large-scale, ordered magnetic fields needed for MHD driven outflows. When a stellar dipole is used as a seed, the field that grows in the disc is anti-parallel to it (see vRB04). This field configuration allows for less efficient mass loading (Dyda15) and we find outflows are $\sim 10$ weaker than when a disc seed field is used. Some wind models, such as the X-wind model (see Shu et al 1994) require a parallel dipole/disc field configuration. Studying the magnetic field topology generated by dynamos is thus a powerful tool for discriminating between viable magnetic wind scenarios. Outflow properties may help us determine whether fields were advected their magnetic field from the ISM and those that generated them local via a disc dynamo mechanism.

The magnetic field topolgy is highly sensitive to the magnetic diffusivity. As we showed, increasing the Prandtl number causes the polarity of the late time dipole moment to change. Physically these findings can be reconciled by understanding that three time scales are at play - the dynamo, the viscous and diffusive time scales. If the dynamo time scale dominates, the viscous disc becomes magnetically dominated and breaks up. This is the regime described in Stepinski \& Levy (1988) who concluded that strong local dynamo modes could disrupt the global modes of the disc. If the diffusive time scale dominates then the field does not grow sufficiently to launch outflows. When time scales are comparable we are in the regime explored in our simulations where outflows can be enhanced but the overall disc dynamics are not disrupted. The relative strength of the viscous and diffusive effects determines the sign of the late time dipole field and the field topology.

Our treatment is limited in that we assumed homogenous values for $\alpha$ and our subgrid dynamo model assumed the simplest possible treatment, namely isotropic turbulence. We treated viscosity, diffusivity and the $\alpha\Omega-$dynamo as three separate effects. However, if they are due to the same microphysics, small scale turbulence, it may actually be unphysical to vary each $\alpha$ parameter independently. Disc accretion studies have already moved away from $\alpha-$disc models, as they have found that the Shakura-Sunyaev picture only holds when coarse graining over the disc in time. This problem requires treating the viscosity self-consistently by resolving the MRI. The $\alpha \Omega-$dynamo effects may then be modeled self-consistently, eliminating the need to consider potentially unphysical parts of the $\alpha_d-\alpha_{\nu}-\alpha_{\eta}$ phase space.

\section*{Acknowledgments}

Resources supporting this work were provided by the NASA High-End
Computing (HEC) Program through the NASA Advanced Supercomputing
(NAS) Division at the NASA Ames Research Center and the NASA
Center for Computational Sciences (NCCS) at Goddard Space Flight
Center. S.D and R.V.E.L were supported by NASA grant NNX14AP30G
and NSF grant AST-1211318. A.V.K acknowledges the support from the Russian Academic
Excellence Projects ``5top100". I.W acknowledges support from NASA grant
NNX13AH42G. 

\appendix

\section{Numerically Extracting Multipole Moments}
\label{sec:multipoles}
Any magnetic field can be described by its multipole expansion. Below, we derive formulae for the axysymmetric dipole, quadrupole and octupole components of a magnetic multipole expansion in terms of surface integrals in cylindrical coordinates. These are used to numerically extract the multipole coefficients of the magnetic field configurations in our simulations.

\begin{figure*}
                \centering
                \includegraphics[width=\textwidth]{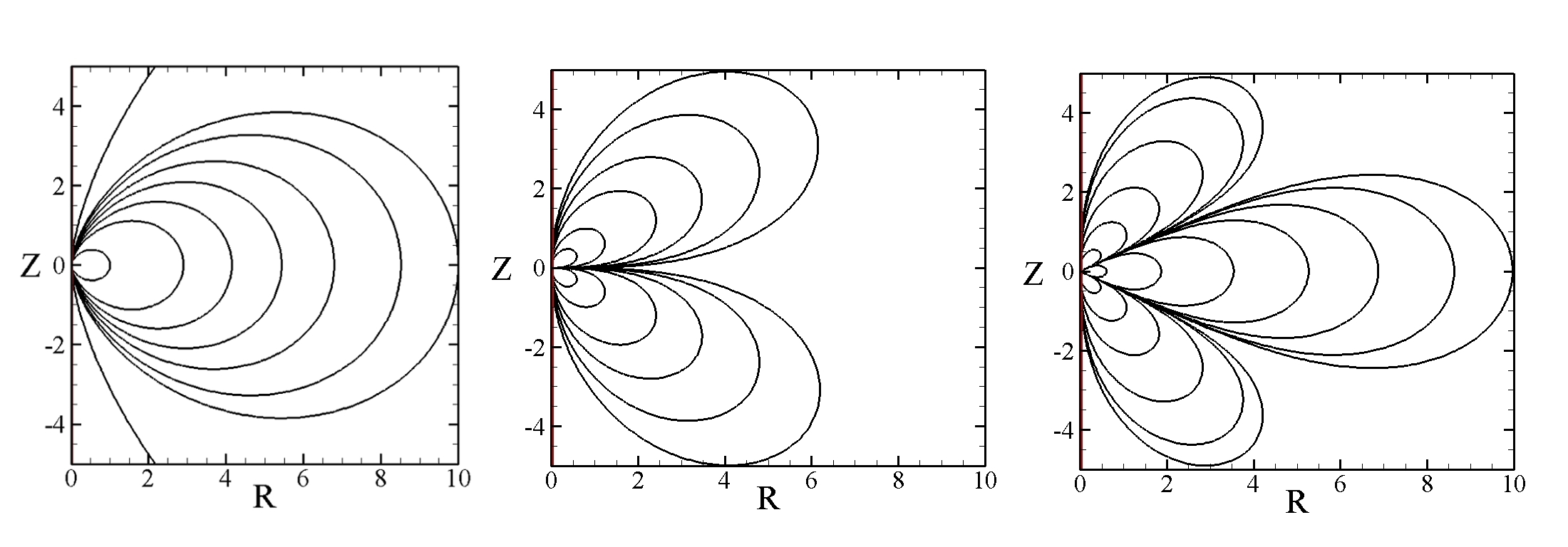}
        \caption{Flux lines of axially symmetric dipole (left), quadrupole (center) and octupole (right).}
\label{fig:multipole}
\end{figure*} 

\subsection{Dipole Moment}
\label{sec:dipole}
Consider a dipole moment, symmetric about the Z-axis with flux function given by
\begin{equation}
\Psi = \mu_{\rm{dip}} \frac{R^2}{\left( R^2 + Z^2 \right)^{3/2}},
\end{equation} 
and poloidal magnetic fields
\begin{equation}
B_R = 3 \mu_{\rm{dip}} \frac{RZ}{\left( R^2 + Z^2 \right)^{5/2}},
\end{equation}
\begin{equation}
B_Z = \mu_{\rm{dip}} \frac{(2Z^2 - R^2)}{\left( R^2 + Z^2 \right)^{5/2}}.
\end{equation}
Given a general magnetic field, we want to extract its dipole moment from its multipole expansion at any given time. In general the dipole moment is given by
\begin{equation}
\mathbf{m} = \frac{1}{2} \int d\mathbf{V} \ \Big[\mathbf{r} \times \mathbf{J} \Big] = \frac{c}{8 \pi} \int d\mathbf{V} \ \Big[ \mathbf{r} \times \left( \nabla \times \mathbf{B} \right) \Big],
\end{equation}
where we work in CGS units and have used Ampere's law. We will integrate over a cylinder $0 < R < a$, $-h < Z < h$, $0 < \phi < 2 \pi$ enclosing the star. We also assume the boundary condition $\Psi(0,Z) = 0$ Using vector identities we may write 
\begin{equation}
\mathbf{m} = \frac{c}{8 \pi} \int d\mathbf{V} \ \Big[ \nabla \left( \mathbf{r} \cdot \mathbf{B} \right) - \left( \mathbf{r} \cdot \nabla \right) \mathbf{B} - \left( \mathbf{B} \cdot \nabla \right) \mathbf{r} \Big]. 
\end{equation}
We are interested in the $\hat{Z}$ component of the magnetic moment, $\mu_{\rm{dip}}$. We calculate the contribution to $\mu_{\rm{dip}}$ from each of the above three terms below.
\begin{equation}
\begin{split}
T_1 \equiv \int d\mathbf{V} \ \nabla_Z \left( \mathbf{r} \cdot \mathbf{B} \right) &= 
\int d\mathbf{S}_Z \left( \mathbf{r} \cdot \mathbf{B} \right) \\ &= 2 \pi \int_0^a R dR \Big. \left( \mathbf{r} \cdot \mathbf{B} \right)\Big|_{-h}^{h},
\end{split}  
\end{equation}
where we have used the divergence theorem to convert the volume integral to a surface integral over the sides of the cylinder and only the top and bottom contribute to the Z-component. Using the definition of magnetic field in terms of the flux function we may write
\begin{equation}
T_1 = 2 \pi \int_0^a  dR \ \Big. R^2 B_R \Big|_{-h}^{h} + 2 \pi h \Big[ \Psi(a,h) + \Psi(a,-h) \Big].   
\end{equation}
The Z-component contribution from the second term is
\begin{equation}
\begin{split}
T_2 &= - \int d\mathbf{V} \ \Big[ \left( \mathbf{r} \cdot \nabla \right) \mathbf{B} \Big]_Z \\ &= - 2 \pi \int R dR \ dZ \left( R \frac{\partial}{\partial R} + Z \frac{\partial}{\partial Z} \right)B_Z.
\end{split}
\end{equation}
Integrating the first term by parts and using the definition of the magnetic flux function we may write
\begin{equation}
T_2 = -2 \pi \int_{-h}^{h}dZ \Big[ a^2 B_Z(a,Z) - 2 \Psi(a,Z) - RZ B_R(a,Z)\Big].
\end{equation}
The third term can be dealt with by rewriting it in terms of the flux function and explicitly carrying out the integration in R to yield
\begin{equation}
T_3 = - 2 \pi \int_{-h}^{h} dZ \ \Psi(a,Z).
\end{equation}
Combining these terms we arrive at the result
\begin{equation}
\begin{split}
\mu_{\rm{dip}} = &\frac{c}{4} \Bigg\{ \int_0^a dR \Big. \ R^2 B_R(R,Z)  \Big|_{-h}^{h} + h \Big[ \Psi(a,h) + \Psi(a,-h) \Big] \\ &- \int_{-h}^{h}dZ \Big[ a^2 B_Z(a,Z) - \Psi(a,Z) - aZ B_R(a,Z)\Big] \Bigg\}, 
\end{split}
\end{equation}
which is easy to implement in our cylindrical code.

\subsection{Quadrupole Moment Calculation}
\label{sec:quadrupole}
\label{ref:octupole}
Consider a quadrupole moment, symmetric about the Z-axis with flux function given by
\begin{equation}
\Psi = \frac{3 \mu_{\rm{quad}}}{4}\frac{R^2 Z}{\left( R^2 + Z^2 \right)^{5/2}},
\end{equation} 
and poloidal magnetic fields
\begin{equation}
B_R = -\frac{3 \mu_{\rm{quad}}}{4} \frac{R (R^2 - 4Z^2)}{\left( R^2 + Z^2 \right)^{7/2}},
\end{equation}
\begin{equation}
B_Z = \frac{3\mu_{\rm{quad}}}{4} \frac{(Z^2 - 3R^2)}{\left( R^2 + Z^2 \right)^{7/2}}.
\end{equation}
Its quadrupole moments are $D_{11} = D_{22} = -D_{33}/2$ where
\begin{equation}
\begin{split}
\mu_{\rm{quad}} \equiv D_{33} =&  \int d\mathbf{V} \ \left[ \mathbf{r} \cdot (\nabla \times \mathbf{J}) \right] Z^2 \\  =& \ 2 \int d\mathbf{V} \ Z \ (\mathbf{r} \times \mathbf{J})_Z
\end{split}
\end{equation}
where as with the dipole, we are interested in the Z-component of $\mu_{\rm{quad}}$. Using Maxwell's equations we can write  
\begin{equation}
\mu_{\rm{quad}} = \frac{c}{2 \pi} \int d\mathbf{V} Z \Big[ \mathbf{r} \times \left( \nabla \times \mathbf{B} \right) \Big]_{Z}.
\end{equation} 
As with the dipole case, this expression can be expanded using a vector identity as
\begin{equation}
\mu_{\rm{quad}} = \frac{c}{2 \pi} \int d\mathbf{V} \ Z \Big[ \nabla \left( \mathbf{r} \cdot \mathbf{B } \right) - \left( \mathbf{r} \cdot \nabla \right) \mathbf{B} - \left( \mathbf{B} \cdot \nabla \right) \mathbf{r} \Big]_{Z}.
\end{equation}
As in the dipole case we calculate each of these terms separately. We are interested in the $\hat{Z}$ component of the quadrupole moment.
\begin{equation}
\begin{split}
T_1 \equiv& \int d\mathbf{V} \ Z \ \nabla_Z \left( \mathbf{r} \cdot \mathbf{B}\right) \\=& - \int d\mathbf{V} \left( \mathbf{r} \cdot \mathbf{B}\right) \hat{z} + Z \left( \mathbf{r} \cdot \mathbf{B}\right) \Big|_{-h}^{h}, 
\end{split}
\end{equation}
where we have integrated by parts.  Writing out the magnetic field in term of the flux function, we expand the first term and carry out some of the intergrals. The second term we evaluate on the top and bottom of the cylinder which yields
\begin{equation}
\begin{split}
T_1 = & \ 2 \pi \int_0^a dR \ R \ \Psi(R,z) \Big|_{h}^{h} - 2 \pi \int_{-h}^{h} dZ \ Z\ \Psi(a,Z) \\ &+ 2 \pi \int_0^a dR \ R\ Z \ (RB_R + ZB_Z)\Big|_{-h}^{h}.
\end{split}
\end{equation}
Expanding the second term and carrying out the angular integral we find
\begin{equation}
\begin{split}
T_2 \equiv& - \int d\mathbf{V} Z (\mathbf{r} \cdot \nabla) \mathbf{B} \\ =& -2 \pi \int \ dR \ dZ \left( ZR^2 \frac{\partial B_z}{\partial R} + Z^2 R \frac{\partial B_z}{\partial Z}\right).
\end{split}
\end{equation}
Performing integration by parts for the R integral on the first term and then writing it out in terms of the flux function allows the integral in R to be evaluated, yielding
\begin{equation}
\begin{split}
T_2 = 2\pi \int_{-h}^{h} dZ \ 2 Z \ & \Psi(a,Z) - \ a^2 Z B_z(a,Z) \\ &+ 2\pi \int_{-h}^{h} dZ a Z^2 B_R(a,Z). 
\end{split}
\end{equation}
For the final term, we take the divergence, expand it in terms of the flux function, carry out the integral in R and find
\begin{equation}
T_3 = - \int d\mathbf{V} Z (\mathbf{B} \cdot \nabla) Z = - 2 \pi \int_{-h}^{h} dZ \ Z \ \Psi(a,Z).
\end{equation} 
Combining these terms we arrive at the result
\begin{equation}
\begin{split}
\mu_{\rm{quad}} = c & \int_0^a dR \Big[ R \Psi(R,Z) + RZ(RB_R + ZB_Z) \Big] \Bigg|_{-h}^{h} \\ &+  c\int_{-h}^{h} dZ \Big[ a Z^2 B_R(a,Z)  -  a^2 Z B_z(a,Z) \Big].
\end{split}
\end{equation}

\subsection{Octupole Moment Calculation}
\label{sec:octupole}
Consider an octupole moment, symmetric about the Z-axis with flux function given by
\begin{equation}
\Psi = \frac{\mu_{\rm{oct}}}{2}\frac{R^2 (4Z^2 - R^2)}{\left( R^2 + Z^2 \right)^{7/2}},
\end{equation} 
and poloidal magnetic fields
\begin{equation}
B_R = \frac{5 \mu_{\rm{oct}}}{2} \frac{RZ (4Z^2 - 3R^2)}{\left( R^2 + Z^2 \right)^{9/2}},
\end{equation}
\begin{equation}
B_Z = \frac{\mu_{\rm{oct}}}{2} \frac{(8Z^4 - 24Z^2R^2 + 3R^4)}{\left( R^2 + Z^2 \right)^{9/2}}.
\end{equation}
The octupole moment is given by
\begin{equation}
\mu_{\rm{oct}} \equiv 2 \sqrt{\frac{\pi}{7}}M_{30},
\end{equation}
where 
\begin{equation}
M_{30} = \frac{1}{4} \int d\mathbf{V} \left[ \mathbf{r} \times \mathbf{J}\right] \cdot \nabla \left( r^3 Y_{30} \right),
\end{equation}
and
\begin{equation}
r^3 Y_{30} = \sqrt{\frac{7}{4\pi}} r^3 P_3(\cos \theta) = \frac{1}{4}\sqrt{\frac{7}{\pi}} Z \left( 2 Z^2 - 3 R^2 \right),
\end{equation}
is a spherical harmonic that we have written in terms of the Legendre polynomial $P_3$. Expanding the cross product and using Amperes law to write out the current in terms of the magnetic field and evaluating the azimuthal integral we find
\begin{equation}
\begin{split}
\mu_{\rm{oct}} = \frac{3c}{16} \Bigg[ 4 \int dR dZ R^2 Z^2 \left( \frac{\partial B_R}{\partial Z} - \frac{\partial B_Z}{\partial R} \right)
\\ - \int dR dZ R^4 \left( \frac{\partial B_R}{\partial Z} - \frac{\partial B_Z}{\partial R} \right) \Bigg]. 
\end{split}
\end{equation}
Evaluate each of these terms separately. The first term is
\begin{equation}
\begin{split}
T_1 \equiv & \int dR dZ R^2 Z^2 \left( \frac{\partial B_R}{\partial Z} - \frac{\partial B_Z}{\partial R} \right) \\
=& \int dR R^2 h^2 B_R\Bigg. \Bigg|_{-h}^{h} - 2 \int dR \ dZ R \Psi + 2 \int dR RZ \Psi \Bigg. \Bigg|_{-h}^{h} \\ -& \int dZ Z^2 a^2 B_Z(a,Z) + 2 \int dZ Z^2 \Psi(a,Z),
\end{split}
\end{equation}
where we have used integration by parts and the definition of magnetic field in terms of the magnetic flux function. Similarly, the second term is
\begin{equation}
\begin{split}
T_2 \equiv & \int dR dZ R^4 \left( \frac{\partial B_R}{\partial Z}  + \frac{\partial B_Z}{\partial R} \right) \\ =& \int dR R^4 B_R \Bigg. \Bigg|_{-h}^{h} -\int dZ a^4 B_Z(a,Z) \\ &- 8 \int dR dZ R \Psi + 4 \int dZ a^2 \Psi(a,Z).  
\end{split}
\end{equation}
Combining these terms we find
\begin{equation}
\begin{split}
\mu_{\rm{oct}} =&  \frac{3c}{16} \Bigg[ 4 \int dR \ R^2h^2 B_{R}\Big. \Big|_{-h}^{h} + 8 \int dR \ R Z \Psi \Big. \Big|_{-h}^{h} \\ &- 4 \int dZ \ Z^2a^2 B_{Z}(a,Z) + 8 \int dZ \ Z^2 \Psi(a,Z) \\ &- \int dR \ R^4 B_{R}\Big. \Big|_{-h}^{h} + \int dZ a^4 B_Z(a,Z) \\ 
&- 4 \int dZ \ a^2 \Psi(a,Z) \Bigg].
\end{split}
\end{equation}
We note the cancellation of the volume integrals in $T_1$ and $T_2$, allowing the final expression to again only depend on surface integrals.

\label{lastpage}

\end{document}